\documentstyle[aps,prl,multicol,fancyheadings,epsfig,amsbsy,amssymb,amstex]{revtex}

\def\bbbc{{\mathchoice {\setbox0=\hbox{$\displaystyle\rm C$}\hbox{\hbox 
to0pt{\kern0.4\wd0\vrule height0.9\ht0\hss}\box0}} 
{\setbox0=\hbox{$\textstyle\rm C$}\hbox{\hbox 
to0pt{\kern0.4\wd0\vrule height0.9\ht0\hss}\box0}} 
{\setbox0=\hbox{$\scriptstyle\rm C$}\hbox{\hbox 
to0pt{\kern0.4\wd0\vrule height0.9\ht0\hss}\box0}} 
{\setbox0=\hbox{$\scriptscriptstyle\rm C$}\hbox{\hbox 
to0pt{\kern0.4\wd0\vrule height0.9\ht0\hss}\box0}}}} 

\begin{document} 
\draft 
\title{Ferromagnetism in the Two-Dimensional Periodic Anderson Model} 
\author{C. D. Batista,$^1$ J. Bon\v ca,$^2$ and J. E. Gubernatis$^1$} 
\address{$^1$Center for Nonlinear Studies and Theoretical Division\\ 
Los Alamos National Laboratory, Los Alamos, NM 87545\\ 
$^2$ Department of Physics, FMF\\ University of Ljubljana and J. Stefan  
Institute, Ljubljana, Slovenia} 
\date{\today} 
\maketitle 
\begin{abstract} 
Using the constrained-path Monte Carlo method, we studied the magnetic 
properties of the two-dimensional periodic Anderson model for electron 
fillings between $1/4$ and $1/2$. We also derived two effective low 
energy theories to assist in interpreting the numerical results. For 
$1/4$ filling we found that the system can be a Mott or a charge 
transfer insulator, depending on the relative values of the Coulomb 
interaction and the charge transfer gap between the two 
non-interacting bands. The insulator may be a paramagnet or 
antiferromagnet. We concentrated on the effect of electron doping on 
these insulating phases. Upon doping we obtained a 
partially saturated ferromagnetic phase for low concentrations of 
conduction electrons. If the system were a charge transfer insulator, 
we would find that the ferromagnetism is induced by the well-known 
RKKY interaction. However, we found a novel correlated hopping 
mechanism inducing the ferromagnetism in the region where the 
non-doped system is a Mott insulator. Our regions of ferromagnetism 
spanned a much smaller doping range than suggested by recent slave 
boson and dynamical mean field theory calculations, but they were 
consistent with that obtained by density matrix renormalization group 
calculations of the one-dimensional periodic Anderson model. 
 
\end{abstract} 
 
\begin{multicols}{2} 
 
\columnseprule 0pt 
 
\narrowtext 
\section{Introduction} 
Identifying the origin of itinerant ferromagnetism in metals and 
specifying simple models exhibiting it are two of the most intriguing 
and long-standing problems in condensed matter physics. Here we report 
the results of low-energy perturbation theory calculations and 
supporting zero temperature quantum Monte Carlo (QMC) simulations that 
suggest the existence and mechanisms for 
ferromagnetic (FM) ground states in the two-dimensional periodic 
Anderson model. 
 
>From a historical point of view, the one-band, nearest-neighbor  
hopping Hubbard model was one of the first models proposed to describe 
itinerant ferromagnetism; however, the ferromagnetic phase has never 
been found at physical parameter values. The numerical calculations, 
for example, have narrowed the extent of this phase to a small region 
around the Nagaoka point \cite{nagaoka}, that is, the strong 
coupling limit for one hole doped away from 
half-filling. Paradoxically, in two-dimensions (2D), this model 
exhibits anti-ferromagnetism at half-filling and anti-ferromagnetic 
correlations around half-filling at weak and intermediate couplings. 
 
Recently, Guerrero and Noack \cite{Guerrero2} listed several 
possible extensions of the Hubbard model that should enhance 
ferromagnetism: (i) the addition of frustrating hopping terms 
\cite{Vollhardt,Daul,Hlubina}, (ii) the inclusion of more than one 
orbital per unit cell, and (iii) the addition of more general 
nearest-neighbor interactions \cite{Vollhardt,Fazekas}. In fact, a 
number of frustrated models with more general interactions, such as 
the $t-t'$ Hubbard model, and multiband models, such as the periodic 
Anderson model (PAM), have ferromagnetic ground states 
\cite{Guerrero2,Fazekas,Guerrero}. In this paper, we focus on the 
properties of the two-dimensional PAM. 
 
The PAM is often used to describe the essential physics of many 
transition metals and rare-earth and actinide metallic compounds 
including the so-called heavy-fermion systems 
\cite{fulde}. The model includes a band of ``light'' uncorrelated 
electrons hybridized with a band of heavy strongly correlated 
electrons.  Despite intense efforts to determine its phases, only a 
few controlled analytical approximations and numerical calculations 
exist for $D\geq2$. Instead, previous work often studied the 
single-impurity Anderson model and focused on the competition between 
Kondo screening and the direct RKKY coupling between the localized 
spins \cite{Hewson,Doniach}. This competition is present when the 
number of conduction electrons is at least similar to the number of 
singly-occupied low-lying $f$-levels.  In $D\ge2$, much less attention 
has been given to the region of the phase diagram where the density of 
electrons in the conduction band is small; however, it is for this 
case that ferromagnetism has been established in one dimensional systems 
\cite{Guerrero,Sigrist,Sigrist2,Troyer,Tsunetsugu,Honner}. 
 
Using the density matrix renormalization (DMRG) method in the 
one-dimension, Noack and Guerrero \cite{Guerrero}, for example, found 
partially and completely saturated ferromagnetism in the PAM.  They 
considered a parameter regime where the energy of orbital state 
$\epsilon_f$ and the strength of the Coulomb repulsion $U$ were 
adjusted so each orbital had just one electron. The position of the 
orbital energy was below the lower band of the non-interacting 
problem, and one electron per orbital corresponds to a 1/4-filled 
non-interacting problem.  For a sufficiently large value of $U$, the 
model exhibited a ferromagnetic ground state. Beyond an 
interaction-dependent value of the doping and a doping-dependent value 
of $U$, this state disappeared. The ferromagnetic 
phase was a peninsula in a phase diagram that was otherwise a sea of 
paramagnetism except at 1/4 and 1/2 filling where the ground state of 
the PAM was antiferromagnetic. 
 
Ferromagnetism seems to be readily found by mean-filed approximations 
in any 
dimensions \cite{Moller,sbmft2,sbmft3,Thavildar-Zadeh,meyer1,meyer2,meyer3}. 
Using a slave-boson mean-field theory (SBMFT) for the symmetric PAM, 
M\"oller and W\"olfe \cite{Moller} found results similar to those of 
Noack and Guerrero.  At 1/2 
filling, they found a paramagnetic (PM) or antiferromagnetic (AF) 
phase depending on the value of the Coulomb repulsion $U$. By lowering 
the density of electrons from 1/2 filling, they also found a smooth 
crossover from AF to FM order via a spiral phase.  Just before 1/4 
filling, they got a first-order transition from FM to AF order. More 
recently, the SBMFT calcualtions of Doradzi\' nski and Spalek 
\cite{sbmft2,sbmft3} found wide regions of ferromagnetism in the intermediate 
valence regime that surprisingly extended well below 1/4 filling. 
 
In the low temperature dynamical mean-field theory (DMFT) 
calculations, Tahvildar-Zadeh {\it et al.\/} \cite{Thavildar-Zadeh} 
also found a region of ferromagnetism and studied its temperature 
dependence.  At very low temperatures, their ferromagnetic region 
extended over a wide range of electron density and in many cases 
embraced the electron density of 3/8. At 3/8 filling they proposed a specific 
Kondo-induced mechanism for ferromagnetism that has the 
conduction electrons in a spin-polarized charge density-wave 
anti-aligned with the ferromagnetically aligned local moments on the 
valence orbitals. More recently Meyer and 
Nolting\cite{meyer1,meyer2,meyer3} appended perturbation theory to DMFT 
and also predicted ferromagnetism over a broad ramge of electron 
filling extending below 1/4 filling. 
 
Our previous \cite{Bonca} and new QMC results qualitatively agree with the DMRG 
work; however, the phases we find quantitatively and qualitatively 
disagree with those derived from the mean-field approximations. 
Quantitatively, we find ferromagnetism in a narrower doping range than 
the one predicted by the DMFT and SBMFT calculations.  For densities 
between 3/8 and 1/2, QMC predicts a paramagnetic region. whereas 
mean-field theory predicts a ferromagnetic states in part of that 
region. In fact, at a filling of 3/8 where these calculations predict 
ferromagnetism, we find a novel ground state of an entirely different 
symmetry. Instead of ferromagnetism, QMC finds a resonating spin 
density-wave (RSDW) state; that is, the ground state was a linear 
combination of two degenerate spin-density waves characterized by the 
$(\pi,0)$ and $(0,\pi)$ wave vectors. 
 
We remark that the quantitative differences between the DMRG and QMC 
calculations and the DMFT and SBMFT calculations most likely result 
from the expected breakdown of mean-field theory in one and two 
dimensions. Probably the RSDW state was not found because it was not 
sought. On the other hand, trying to understand the mechanism for 
ferromagnetism is more fundamental. It points to the long-standing 
difficulty of building an understanding of the PAM upon the better 
understood single impurity Anderson model (SIAM) or the analogous 
problem of building an understanding of the Kondo lattice model upon 
the better understood single Kondo impurity problem. Nozi\'eres 
addressed this later problem and proposed a reconciliation in terms of 
what he calls ``the exhaustion picture'' \cite{Nozieres,Nozieres2}. 
 
We note that the electron densities near 1/4 filling place the work of 
Noack and Guerrero \cite{Guerrero} in the exhaustion regime. When the 
Coulomb repulsion $U$ associated with the double occupancy of an 
orbital is large (strong coupling), it is often argued that around 1/4 
filling the PAM behaves like a 1/2-filled one-band Hubbard model. In 
two-dimensions this Hubbard model has an 
antiferromagnetic ground-state generated by a superexchange 
interaction $J_{sx}\sim V^2t^4/U^5$ (for the symmetric case) 
where $V$ is the strength of the 
hybridization in the PAM. It has also been established that the Kondo 
lattice model and the Hubbard model for strong coupling are isomorphic 
with particles in one becoming holes in the other \cite{Lacroix}.  The relevance of 
the Hubbard model gains additional support from Nozi\'eres's 
long-standing argument that the dynamics of the screening clouds are 
described by an effective Hubbard model. 
 
Because of these mappings, it is seems consistent to suggest the 
strong-coupling physics found by Noack and Guerrero would map onto the 
domain of a one-dimensional 1/2-filled Hubbard model. The venerable 
theorem of Lieb and Mattis \cite{Lieb}, however, excludes the 
possibility of the Hubbard model (with nearest neighbor hopping) from 
showing ferromagnetism in one-dimension. Their proof relied on the 
obvious ability to order the electrons along the chain. The two-band 
nature of the PAM, however, prevents this ordering by allowing 
processes not possible in the Hubbard model. A similar 
situation would occur for the two-legged Hubbard model, if it in fact 
shows ferromagnetism. We will argue that in the two-dimensional PAM 
these same processes are responsible for the 
ferromagnetism. They are RKKY and Nagaoka-like and are 
excluded in Nozi\'eres's \cite{Nozieres,Nozieres2} picture and 
Tahvildar-Zadeh {\it et al.\/}'s \cite{Thavildar-Zadeh} 
interpretation of it. It is important to remark that ferromagnetism  
in the PAM is obtained for a large region of parameters which 
include realistic values. 
 
We will base our arguments on the predictions of effective 
Hamiltonians generated from the PAM by perturbation theory and the 
results of our QMC simulations.  For the PAM parameters studied, these 
effective Hamiltonians suggest a paramagnetic or anti-ferromagnetic 
state at 1/4 filling, a ferromagnetic region between 1/4 and 3/8 
filling, a RSDW at 3/8 filling, a paramagnetic region between 3/8 and 
1/2 filling, and an antiferromagnetic state at 1/2 filling. We see all 
these features in the QMC simulations.  
 
In the Section~II we will define the PAM and sketch our derivation of 
the effective Hamiltonians. In Section~III we will summarize 
our numerical method, noting finite size limitations.  Our results 
will be presented in Section~IV. In the Section~V, the Concluding 
Section, we will give a detailed contrast between our picture and 
select other works.

\section{Models} 
 
The PAM is described by the Hamiltonian 
\begin{eqnarray} 
  H &=& -t\sum_{\langle i,j \rangle,\sigma} (d_{i\sigma}^\dagger 
  d^{}_{j\sigma}+d_{j\sigma}^\dagger d^{}_{i\sigma}) 
  +V\sum_{i,\sigma} (d_{i\sigma}^\dagger 
  f^{}_{i\sigma}+f_{i\sigma}^\dagger d^{}_{i\sigma}) \nonumber \\ & & 
  \quad\quad +\epsilon_f\sum_{i,\sigma}n_{i\sigma}^f +\frac{U}{2} 
  \sum_{i,\sigma}n_{i\sigma}^fn_{i\bar {\sigma}}^f 
\label{eq:pam} 
\end{eqnarray} 
where $d_{i\sigma}^\dagger$ and $f_{i\sigma}^\dagger$ create an 
electron with spin $\sigma$ 
in $d$ and $f$ orbitals at site $i$ in a square lattice, and 
$n^f_{i\sigma}=f^{\dagger}_{i\sigma}f^{}_{i\sigma}$ is the number 
operator for the $f$-electrons of spin ${\sigma}$ at site $i$. Elsewhere 
we will use a similar notation to denote quantities like 
$n^d_{i\sigma}=d^{\dagger}_{i\sigma}d^{}_{i\sigma}$, the number 
operator of $d$-electrons. The lattice has $N$ sites and the hopping 
amplitude $t$ between $d$-orbitals is only to nearest-neighbor (n. n.) 
sites. The hopping amplitude $V$ hybridizes different orbitals on the 
same site. We used periodic boundary conditions. 
 
\begin{figure}[tbp] 
\begin{center}
\epsfig{file=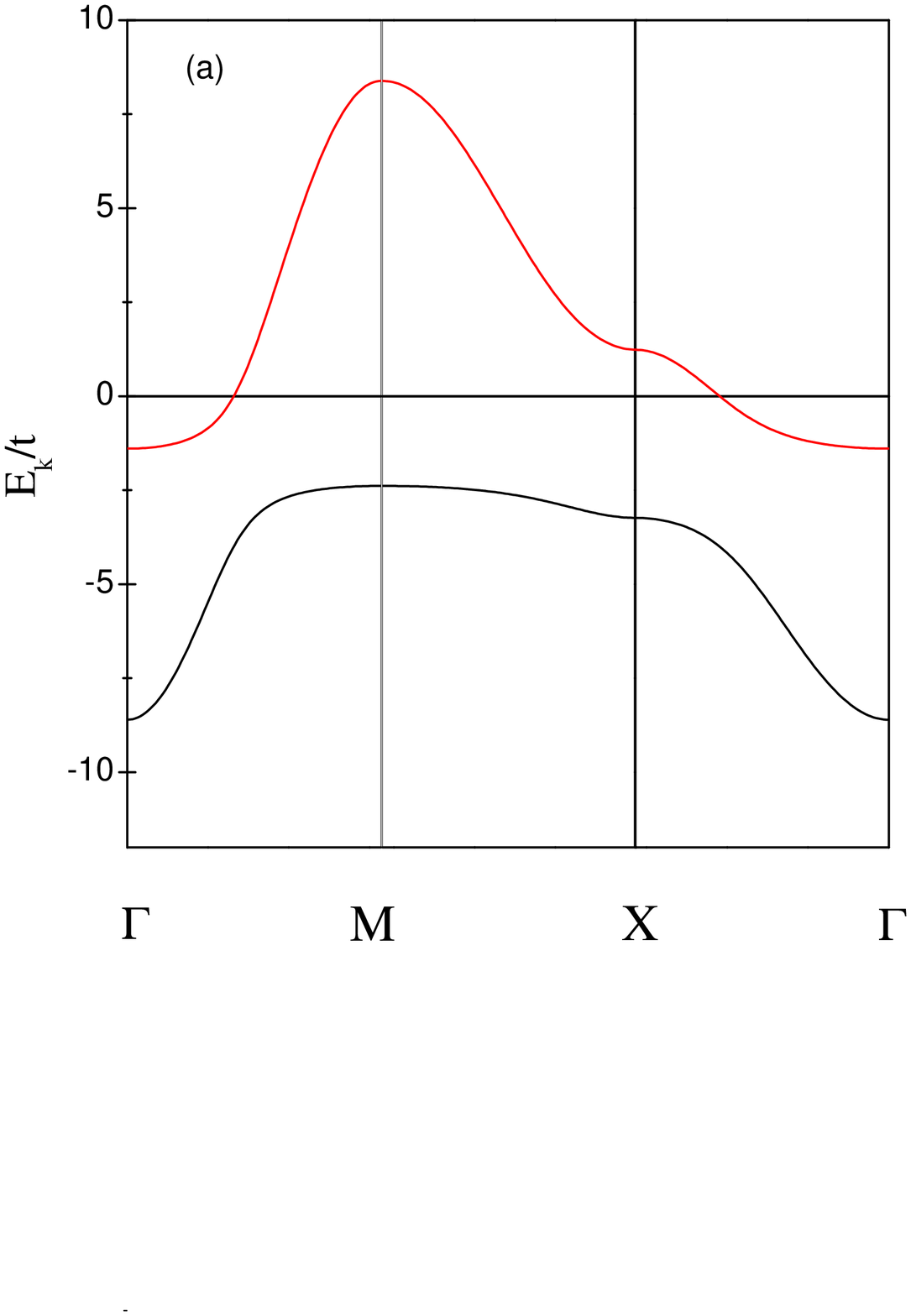,width=70mm,angle=-0}
\epsfig{file=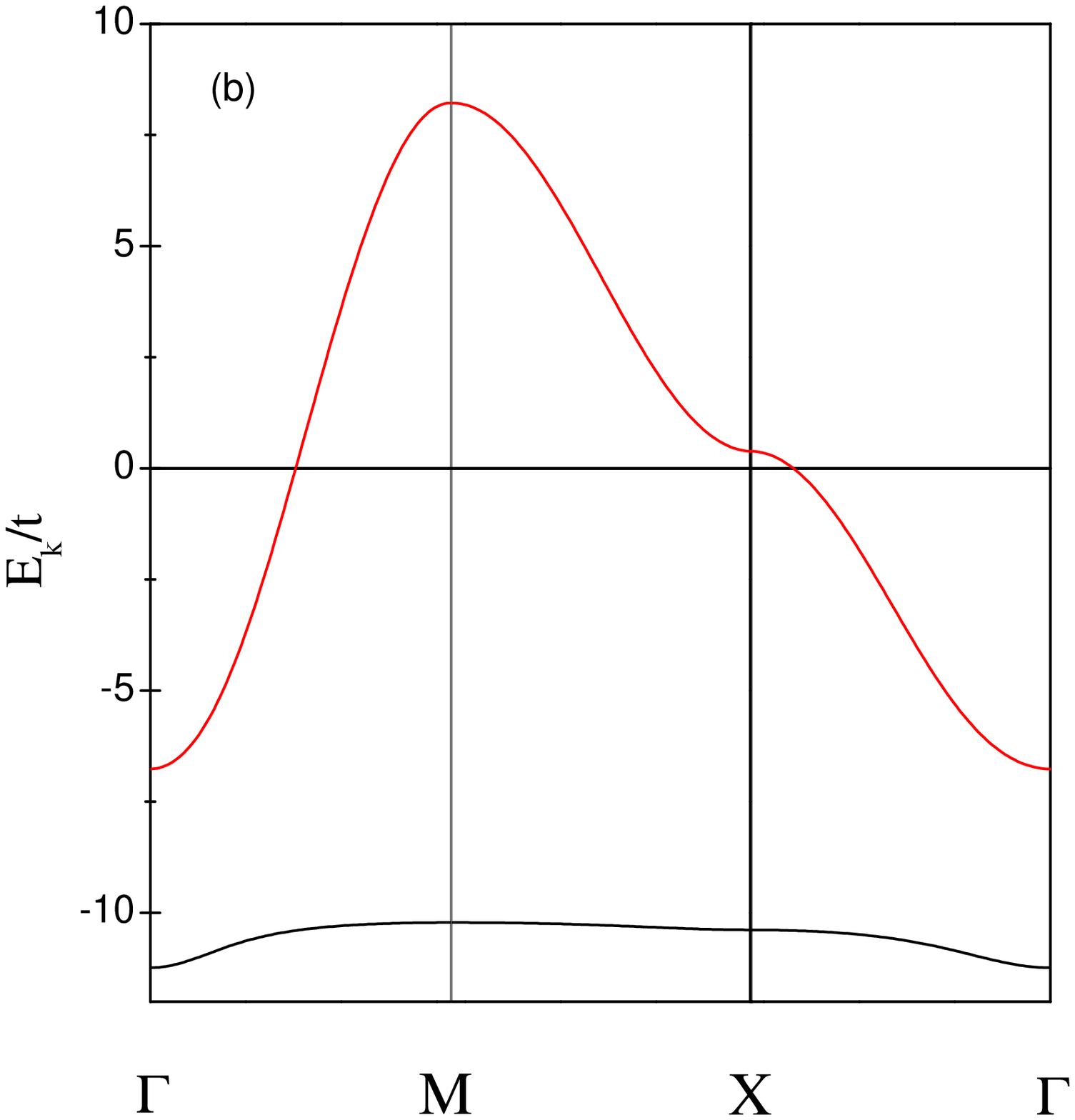,width=70mm,angle=-0}
\end{center}
\caption{Band structure of the non-interacting ($U=0$) two-dimensional 
periodic Anderson model. In units of $t$, $V=-0.5$. In (a) 
$\epsilon_f=0.2$ and (b) $\epsilon_f=0.5$  Ilustrated is the flatnes 
of the lower band for case (b).} 
\label{fig1} 
\end{figure}

>From (\ref{eq:pam}) we define $H_{0}$, the resulting 
Hamiltonian when $U=0$. $H_0$ has two dispersive bands 
\begin{equation} 
 E_\sigma^\pm({\bf k})=\frac{1}{2} \Biggl[ 
   e_{\bf k}+\epsilon_f \pm \sqrt{(e_{\bf k} 
                                            -\epsilon_f)^2+4V^2} 
   \Biggr] 
\end{equation} 
separated by a gap 
\begin{eqnarray} 
 \Delta &=&E_\sigma^{+}(0,0)-E_\sigma^{-}(\pi,\pi)  
        \nonumber \\  
        &=&-4t + \frac{1}{2}\Biggl[ 
        \sqrt{(4t+\epsilon_f)^2+4V^2} + \sqrt{(4t-\epsilon_f)^2+4V^2} 
        \Biggr] 
\end{eqnarray} 
For a square lattice, the energy $e_{\bf k}=-2t(\cos k_x + \cos 
k_y)$. This band structure for $H_0$ is illustrated in 
Fig.~\ref{fig1}. We note if $\epsilon_f$ becomes very negative (doping 
way below the bottom of the lower band), $\Delta$ approaches 
$|\epsilon_f|$.

We also note that the widths of the upper and lower band are 
\begin{eqnarray} 
 W^\pm &=& E_\sigma^\pm(\pi,\pi)-E_\sigma^\pm(0,0) 
\nonumber \\ 
       &=& 4t \pm \frac{1}{2}\Biggl[ 
        \sqrt{(4t-\epsilon_f)^2+4V^2} - \sqrt{(4t+\epsilon_f)^2+4V^2} 
        \Biggr] 
\end{eqnarray}         
As $\epsilon_f$ becomes very negative, $W^-$ approaches zero and $W^+$ 
approaches $W=8t$, the band width when $U=V=0$. 
 
The operators which create quasi-particles in the lower and 
upper bands are of the form 
\begin{eqnarray} 
 \alpha_{{\bf k}\sigma}^\dagger  
    &=&  u^{}_{{\bf k}} f_{{\bf k}\sigma}^\dagger 
                     +v^{}_{{\bf k}} d_{{\bf k}\sigma}^\dagger 
 \nonumber\\ 
 \beta_{{\bf k}\sigma}^\dagger  
    &=&  -v^{}_{{\bf k}} f_{{\bf k}\sigma}^\dagger 
                     +u^{}_{{\bf k}}  d_{{\bf k}\sigma}^\dagger 
\label{alphak} 
\end{eqnarray} 
with 
\begin{eqnarray} 
u_{\bf k}&=& \frac{E^+({\bf k})-\epsilon_f}{\sqrt{(E^+({\bf k})-\epsilon_f)^2+V^2}} 
\nonumber\\ 
v_{\bf k}&=& \frac{-V}{\sqrt{(E^+({\bf k})-\epsilon_f)^2+V^2}} 
\label{ukvk} 
\end{eqnarray} 
 
The symmetric PAM, which has the electron filling $\rho= 1/2$ and 
$U=-\epsilon/2$, has particle-hole symmetry. This symmetry is 
sufficient to prevent the fermion sign problem that plagues QMC 
simulations. Such simulations, performed by Veki\' c et 
al. \cite{vekic}, suggest the existence of a charge and spin gap for 
small values of $U$ with the spin gap disappearing when $U$ is 
increased to some $U_c\sim 2$. Above $U_c$ the system exhibits 
long-range anti-ferromagnetic order. 
 
In a previous work we presented QMC results for the asymmetric 
model. QMC simulations for the asymmetric model experience a sign 
problem which is the reason why we used the constrained-path Monte 
Carlo method \cite{zhang1}. For fixed values of $\epsilon_f$, we varied $U$ and 
hole-doped away from half-filling down to a filling of 3/8. For a 
large enough value of $U$ we also found anti-ferromagnetism at 1/2 
filling. This state was rapidly suppressed upon doping. At 3/8 filling 
we saw a sharp peak appear in the spin-spin correlation at the 
wavenumbers ${\bf k}=(0,\pi)$ and $(0,\pi)$. We interpreted this peak 
as consequence of a state resonating between two degenerate spin-density 
waves characterized by reciprocal wavevectors $(0,\pi)$ and $(0,\pi)$. 
 
In the present work we explore the doping range from 1/4 to 3/8 
filling, arguing for a region of ferromagnetism. Part of our arguments 
will be based on the properties of effective Hamiltonians for two 
different regions of parameters used in the simulations. These 
Hamiltonians will be derived in Section~IV.  We now summarize the 
the constrained-path Monte Carlo method.

\section{Numerical Method} 
 
The constrained path Monte Carlo (CPMC) method is extensively 
described and benchmarked elsewhere \cite{zhang1}. Here we only 
discuss its basic strategy and approximation.  In the CPMC method, the 
ground-state wave function $|\psi_0\rangle$ is projected from a known 
initial wave function $|\psi_T\rangle$ by a branching random walk in 
an over-complete space of Slater determinants $|\phi\rangle$.  In such 
a space, we can write $|\psi_0\rangle = \sum_\phi \chi(\phi) 
|\phi\rangle$.  The random walk produces an ensemble of 
$|\phi\rangle$, called random walkers, which represent 
$|\psi_0\rangle$ in the sense that their distribution is a Monte Carlo 
sampling of $\chi(\phi)$, that is, a sampling of the ground-state wave 
function. 
 
More specifically, starting with some trial state $|\psi_T\rangle$, we 
project out the ground state by iterating 
\begin{equation} 
 |\psi'\rangle = e^{-\Delta\tau (H-E_T)}|\psi\rangle 
\end{equation} 
where $E_T$ is some guess of the ground-state energy.  Purposely 
$\Delta\tau$ is a small parameter so for $H=T+V$ we can write 
\begin{equation} 
 e^{-\Delta\tau H}\approx e^{-\Delta\tau T/2} 
	                  e^{-\Delta\tau V} 
                          e^{-\Delta\tau T/2} 
\end{equation} 
where $T$ and $V$ are the kinetic and potential energies.  
 
For the study at hand, the initial state $|\psi_T\rangle$ is the direct 
product of two spin Slater determinants, i.e., 
\begin{equation} 
 |\psi_T\rangle = \prod_\sigma |\phi_T^\sigma\rangle 
\end{equation} 
Because the kinetic energy is a quadratic form in the creation and 
destruction operators for each spin, the action of its exponential on 
the trial state is simply to transform one direct product of Slater 
determinants into another. While the potential energy is not a 
quadratic form in the creation and destruction operators, its 
exponential is replaced  by sum of exponentials of such forms via 
the discrete Hubbard-Stratonovich transformation 
\begin{eqnarray*} 
\lefteqn{e^{-\Delta\tau U n_{i,\sigma}n_{i,-\sigma}}}\\ 
    & & \hspace{1cm} = \frac{1}{2} 
   \sum_{x=\pm 1}e^{-x \Delta\tau J(n_{i,\sigma}-n_{i,-\sigma})} 
               e^{\frac{1}{2}\Delta\tau U(n_{i,\sigma}+n_{i,-\sigma})} 
\end{eqnarray*} 
provided $U\ge 0$ and $\cosh \Delta\tau J = e^{-\Delta\tau 
U/2}$. Accordingly we re-express the iteration step as 
\begin{equation} 
 \prod_\sigma |\phi_\sigma'\rangle = \int d\vec x\, P(\vec x) 
    \prod_\sigma B_\sigma(\vec x)|\phi_\sigma\rangle 
\end{equation} 
where $\vec x =(x_1,x_2,\dots,x_N)$ is the set of Hubbard-Stratonovich 
fields (one for each lattice site), $N$ is the number of lattice 
sites, $P(\vec x)=(\frac{1}{2})^N$ is the probability distribution for 
these fields, and $B_\sigma(\vec x)$ is an operator function of these 
fields formed from the product of the exponentials of the kinetic and 
potential energies. 
 
The Monte Carlo method is used to perform the multi-dimensional 
integration over the Hubbard-Stratonovich fields. It does so by 
generating a set of random walkers initialized by replicating 
$|\psi_T\rangle$ many times. Each walker is then propagated independently by 
sampling a $\vec x$ from $P(\vec x)$ and propagating it 
with $B(\vec x)$. After the propagation has ``equilibrated,'' the sum 
over the walkers provides an estimate of the ground-state wave function  
$|\psi_0\rangle$. 
 
In practice we performed an importance-sampled random walk by using 
the transformed iterative equation 
\begin{equation} 
 \prod_\sigma|\phi_\sigma'\rangle = y^{-1}\int d\vec x\, \tilde P(\vec 
x)\prod_\sigma B_\sigma(\vec x)| 
                \prod_\sigma\phi_\sigma\rangle 
\end{equation} 
In this equation 
\begin{equation} 
\tilde P(\vec x) = {\cal Z}P(\vec x) 
    \frac{\prod_\sigma\langle\phi_T^\sigma|B_\sigma(\vec x)|\phi_\sigma\rangle} 
         {\prod_\sigma\langle\phi_T^\sigma|\phi_\sigma\rangle} 
\end{equation} 
Thus importance sampling changes the probability distribution of the 
Hubbard-Stratonovich fields, biasing it towards the generation of 
states with large overlap with the initial state.  The factor ${\cal Z}$ is 
the normalization constant for the new distribution.  It is associated 
with the weight assigned to each walker and, the weight is used in a 
branching process to control the variance of the results. We will not 
discuss this process here. 
 
We used two different estimators for the expectation values of some 
observable ${\cal O}$. One is the mixed estimator 
\begin{equation} 
 \langle {\cal O}\rangle_{\mathrm{mixed}} = 
     \frac{\langle\psi_T|{\cal O}|\psi_0\rangle} 
          {\langle\psi_T|\psi_0\rangle} 
\end{equation} 
and the other is the back-propagated estimator 
\begin{equation} 
 \langle{\cal O}\rangle_{\mathrm{bp}} = 
     \frac{\langle\psi_T|e^{-\ell\Delta\tau H}{\cal O}|\psi_0\rangle} 
          {\langle\psi_T|e^{-\ell\Delta\tau H}|\psi_0\rangle} 
\end{equation} 
where $|\psi_0\rangle$ is the QMC estimate of the ground state and 
$\ell$ is typically in the range of 20 to 40.  For observables that 
commute with the Hamiltonian, the mixed estimator is a very accurate 
one and converges to the exact answer as $|\psi_0\rangle$ converges to 
exact ground state. For observables that do 
not commute with the Hamiltonian, like correlation functions, 
the back-propagated estimator has been found to give very accurate 
estimates of ground-state properties. Significant differences between the 
predictions of these two estimators often exist. 
 
To completely specify the ground-state wave function for a system of 
interacting electrons, only determinants satisfying 
$\langle \psi_0|\phi_\sigma\rangle>0$ are needed because $|\psi_0\rangle$ 
resides in either of two degenerate halves of the Slater determinant 
space, separated by a nodal surface ${\bf N}$ that is defined by 
$\langle \psi_0|\phi_\sigma\rangle = 0$.  The degeneracy is a consequence of 
both $|\psi_0\rangle$ and $-|\psi_0 \rangle$ satisfying Schr\"odinger's 
equation. The sign problem occurs because walkers can cross ${\bf N}$ 
as their orbitals evolve continuously in the random 
walk. Asymptotically they populate the two halves equally, leading to 
an ensemble that has zero overlap with $|\psi_0\rangle$.  If ${\bf N}$ 
were known, we would simply constrain the random walk to one half of 
the space and obtain an exact solution of Schr\"odinger's equation. 
In the constrained-path QMC method, without {\it a priori\/} knowledge 
of ${\bf N}$, we use a trial wave function $|\psi_T\rangle$ and 
require $\langle \psi_T|\phi_\sigma\rangle>0$.  This is what is called the 
constrained-path approximation. 
 
The quality of the calculation clearly depends on the trial wave 
function $|\psi_T\rangle$. Since the constraint only involves the 
overall sign of its overlap with any determinant $|\phi\rangle$, it 
seems reasonable to expect the results to show some insensitivity to 
$|\psi_T\rangle$.  Through extensive benchmarking on the Hubbard 
model, it has been found that simple choices of this function can give 
very good results \cite{zhang1}. 
 
Besides as a starting point and as a condition constraining a random 
walker, we also use $|\psi_T\rangle$ as an importance function. To 
reduce variance, we use $\langle \psi_T|\phi_\sigma\rangle$ to bias the random 
walk into those parts of Slater determinant space that have a large 
overlap with the trial state. For all three uses of $|\psi_T\rangle$, 
it clearly is advantageous to have $|\psi_T\rangle$ approximate 
$|\psi_0\rangle$ as closely as possible. Only in the constraining of 
the path does $|\psi_T\rangle \not= |\psi_0\rangle$ generate an 
approximation. 
 
We constructed $ |\psi_T\rangle = \prod_\sigma |\phi_T^\sigma\rangle$ 
from the eigenstates of the non-interacting problem. Because the 
z-component of the total spin angular momentum is a good quantum 
number, we could choose unequal numbers of up and down electrons to 
produce trial states and hence ground states with 
$S_z=\frac{1}{2}(N_\uparrow-N_\downarrow)$. Whenever possible, we 
would simulate closed shells of up and down electrons, as such cases 
usually provided energy estimates with the least statistical error, 
but because we wanted to study the ground state energy as a function 
of $S_z$, we frequently had to settle for just the up or down shell 
being closed. In some cases, the desired value of $S_z$ could not be 
generated from either shell being closed. Also we would select the 
non-interacting states so $|\psi_T\rangle$ would be translationally 
invariant, even if these states used did not all come from the Fermi 
sea.  The use of unrestricted Hartree-Fock eigenstates to 
generate $|\phi_T^\sigma$ instead of the non-interacting eigenstates 
generally produced no significant improvement in the results.

\section{Results} 
 
\subsection{Effective Hamiltonian: Wannier Orbital Approach} 
 
Our first effective Hamiltonian explicitly targets cases where the 
lower band of the non-interacting model is very flat. Such cases exist 
for $-\epsilon_f\gtrsim W/2 > V>0$.  In this regime of parameters 
and around 1/4 filling, single electron occupancy of the f-states 
can occur because of the depth of the orbital state as opposed to 
the double occupancy penalty of the Coulomb repulsion. We will begin 
by building Wannier orbital operators \cite{Aligia} for each band from the 
quasi-particle operators defined in (\ref{alphak}) 
\begin{eqnarray} 
 \alpha_{j\sigma}^\dagger &=& \sum_l \bigl(a^{}_{jl} 
 f_{l\sigma}^\dagger +b^{}_{jl} d_{l\sigma}^\dagger\bigr) \nonumber\\ 
 \beta_{j\sigma}^\dagger &=& \sum_l \bigl(-b^{}_{jl} 
 f_{l\sigma}^\dagger +a^{}_{jl} d_{l\sigma}^\dagger\bigr)\\ 
 a_{jl} &=&{\frac{1}{N}} \sum_{\bf k} e^{i{\bf k}\cdot{\bf R}_{ij}} u_{\bf k} 
\nonumber\\ 
b_{jl} &=&{\frac{1}{N}} \sum_{\bf k} e^{i{\bf k}\cdot{\bf R}_{ij}} v_{\bf k} 
\label{wannier} 
\end{eqnarray} 
where ${\bf R}_{ij}={\bf r}_i-{\bf r}_j$.  
 
Rewriting $H_0$ in the Wannier basis, we find that 
\begin{equation} 
H_0=\sum_{i,j,\sigma} \bigl(\tau_{ij}^{\alpha} 
\alpha_{i\sigma}^\dagger \alpha^{}_{j\sigma} 
+ \tau_{ij}^{\beta}  
\beta_{i\sigma}^\dagger \beta^{}_{j\sigma}\bigr) 
\end{equation} 
with 
\begin{eqnarray} 
\tau_{ij}^{\alpha} 
  = &-&t\sum_{\langle l,n \rangle} b_{il} b_{jn}+ \epsilon_f \sum_{l} 
    a_{il} a_{jl}  
\nonumber\\ 
&+&V \sum_{l} (b_{il} a_{jl}+b_{jl} a_{il}) 
\nonumber\\ 
\tau_{ij}^{\beta} 
  = &-& t\sum_{\langle l,n \rangle} a_{il} a_{jn}+ \epsilon_f \sum_{l} 
    b_{il} b_{jl}  
\nonumber\\ 
&-&V \sum_{l} (b_{il} a_{jl}+b_{jl} a_{il}) 
\end{eqnarray} 
$H_0$ is simply the sum of two hopping terms corresponding to the 
lower and the upper bands. By construction, no hybridization exists 
between these two bands. The cost for this simplification is non-zero 
hoppings $\tau_{ij}^{\alpha}$ and $\tau_{ij}^{\beta}$ between any pair 
of Wannier orbitals $i$ and $j$ in the same band. 
   
Next we rewrite the interaction term  
\begin{equation} 
H_U=\frac{1}{2} U 
\sum_{j,\sigma}n_{j\sigma}^fn_{j\bar \sigma}^f 
\end{equation}  
in the Wannier basis 
\begin{eqnarray} 
H_U=U\sum_{j,i,i',l,l'} (a^{}_{ij} \alpha_{i\uparrow}^\dagger 
&+&b^{}_{ij} \beta_{i\uparrow}^\dagger) (a^{}_{i'j} 
\alpha^{}_{i'\uparrow}+b^{}_{i'j} \beta^{}_{i'\uparrow}) 
\nonumber\\ 
\times(a^{}_{lj} \alpha_{l\downarrow}^\dagger &+& b^{}_{lj} 
\beta_{l\downarrow}^\dagger) (a^{}_{l'j} 
\alpha^{}_{l'\downarrow}+b^{}_{l'j} \beta^{}_{l'\downarrow}) 
\label{hu} 
\end{eqnarray} 
This expression appears more complex than the one in the 
original basis; however, from it we can more conveniently derive a 
low energy effective Hamiltonian for electron fillings less than 1/2 
filling.

To do this we first require that $\Delta > U$, i.e. the system is a Mott 
insulator for 1/4 filling,  so we can initially 
consider a $H_{1\, band}^{(0)}$ that does not have processes involving the 
upper band 
\begin{equation} 
H_{1\, band}^{(0)}=\sum_{i,j,\sigma} \tau_{ij}^{\alpha} 
\alpha_{i\sigma}^\dagger \alpha^{}_{j\sigma}+  
U \sum_{i,i',l,l'}g^{}_{ii'll'} 
\alpha_{i\uparrow}^\dagger  \alpha^{}_{i'\uparrow} 
\alpha_{l\downarrow}^\dagger \alpha^{}_{l'\downarrow} 
\end{equation} 
with $g_{ii'll'}=\sum_j a_{ij}a_{i'j}a_{lj}a_{l'j}$. To identify more 
easily the physically different contributions, we rewrite the $U$ term 
to produce 
\begin{eqnarray} 
H_{1\, band}^{(0)}&=&\sum_{i,j,\sigma} \tau_{ij}^{\alpha} 
\alpha_{i\sigma}^\dagger \alpha^{}_{j\sigma}+ 
\tilde U \sum_i    n^{\alpha}_{i\uparrow}    n^{\alpha}_{i\downarrow}  
 \nonumber\\                                          
&+&\sum_{i,l,i\neq l} J^e_{il} \biggl({\bf S}_i\cdot {\bf S}_l 
                   - \frac {n^{\alpha}_i n^{\alpha}_l}{4}\biggr) 
\nonumber\\ 
&+& U \sum_{i,l,l',l\neq l'} \omega^{}_{ill'}(n^{\alpha}_{i\uparrow} 
\alpha_{l\downarrow}^\dagger \alpha^{}_{l'\downarrow} 
+n^{\alpha}_{i\downarrow} \alpha_{l\uparrow}^\dagger \alpha^{}_{l'\uparrow}) 
\nonumber\\ 
&+&U \sum'_{i,i',l,l'} g^{}_{ii'll'} 
\alpha_{i\uparrow}^\dagger  \alpha^{}_{i'\uparrow} 
\alpha_{l\downarrow}^\dagger  \alpha^{}_{l'\downarrow} 
\label{H01band} 
\end{eqnarray} 
with 
\begin{eqnarray} 
\tilde U&=&{\frac {U}{N}} \sum_{i,j} a_{ij}^4 
\nonumber\\ 
J^e_{il}&=&2U\sum_j a_{ij}^2 a_{lj}^2 
\nonumber\\ 
\omega_{ill'}&=& \sum_j a_{ij}^2 a_{jl} a_{jl'} 
\end{eqnarray} 
where $\sum'$ means that there are no repeated indices.  
We see that $H_{1\, band}^{(0)}$ is an extended Hubbard model with long 
range hoppings, a ferromagnetic exchange interaction, correlated 
hoppings, and a term which destroys a spin-anti-aligned pair of 
electrons in sites $i'$ and $l'$ and creates an anti-aligned pair at 
$i$ and $l$.   
 
Again for a large range of parameters, the lower band of the PAM is 
quite flat. If we regard $|\epsilon_f|$ as very large, we can Taylor 
series expand Eq.~\ref{ukvk}, substitute the result into 
Eq.~\ref{wannier}, and obtain 
\begin{eqnarray} 
  a_{ij} &\approx& \left\{\begin{array}{ll} 
                          \delta_{ij}, & {\rm for}\ i=j \\ 
                         -tV^2/|\epsilon_f|^3, & \mbox{for  
$i$ and $j$ n. n.)} 
                           \end{array} 
                   \right.\nonumber\\ 
  b_{ij} &\approx& \left\{\begin{array}{ll} 
                          -V/|\epsilon_f|,        & {\rm for}\ i=j \\ 
                          -tV/{|\epsilon_f|}^2, & \mbox{for $i$ and 
$j$ n. n.} 
                           \end{array} 
                   \right.\nonumber\\ 
\end{eqnarray} 
Matrix elements for $i$ and $j$ beyond nearest neighbors (n. n.) are 
smaller by higher powers of $V/|\epsilon_f|$. Thus the Wannier 
operator $\alpha_j^\dagger$ is predominately $f_j^\dagger$ as the 
amplitudes $a_{ij}$ and $b_{ij}$ strongly decrease with the distance 
between $i$ and $j$. 
 
With these results we see that $J^e_{il}$ and $\omega_{ill'}$ are  
poportional to $t^2V^4/|\epsilon_f|^6$, while $g_{ii'll'}$ is proportional 
to $t^3V^6/|\epsilon_f|^9$. This means that we can neglect the last term of 
$H_{1\, band}^{(0)}$, 
\begin{eqnarray} 
H_{1\, band}^{(0)}&\approx&\sum_{i,j,\sigma} \tau_{ij}^{\alpha} 
\alpha_{i\sigma}^\dagger \alpha^{}_{j\sigma}+ 
\tilde U \sum_i    n^{\alpha}_{i\uparrow}    n^{\alpha}_{i\downarrow}  
 \nonumber\\                                          
&+&\sum_{i,l,i\neq l} J^e_{il} \biggl({\bf S}_i\cdot {\bf S}_l 
                   - \frac {n^{\alpha}_i n^{\alpha}_l}{4}\biggr)  
\nonumber\\ 
&+& U \sum_{i,l,l',l\neq l'} \omega^{}_{ill'}(n^{\alpha}_{i\uparrow} 
\alpha_{l\downarrow}^\dagger \alpha^{}_{l'\downarrow} 
+n^{\alpha}_{i\downarrow} \alpha_{l\uparrow}^\dagger \alpha^{}_{l'\uparrow}) 
\label{H1bd} 
\end{eqnarray} 
 
In this one band Hamiltonian the Coulomb repulsion is no longer just 
on-site. Its spatial extension depends on the spatial extension of the 
Wannier orbitals.  Because of original form of this interaction, it 
still only affects singlet states, but these states can now be 
non-local. We note that even if $U/t\sim 1$ this one-band model is in 
the strong-coupling regime: $\bar U\sim U$ and $\tau_{ij}^{\alpha}\approx 
t(V^2/|\epsilon_f|^2)$ if $i$ and $j$ are nearest neighbors so $\bar 
U/\tau_{ij}^{\alpha} \gg 1$. In the new basis the narrow band appears 
narrower, the heavy fermions appear heavier, but the interaction 
experiences little renormalization. 
 
The non-locality is the origin of the ferromagnetic Heisenberg 
term. In this term, provided $U$ is not arbitrarily large, the 
ferromagnetic exchange interaction $J^e_{il}\sim Ut^2V^4/|\epsilon_f|^6$, 
however, is smaller than the antiferromagnetic super-exchange 
interaction which is of order $(\tau_{il}^{\alpha})^2/\tilde U \sim 
t^2V^4/U|\epsilon_f|^4$. we  
remark is that the magnitude of the AF interaction is very small. 
For this reason, the lowest order terms involving 
the upper band are crucial to determine the magnetic phase of the system doped 
above 1/4 filling.  
         
These lowest order upper band processes come from terms in the 
Eq.~$\ref{hu}$  
with one $\beta$ operator and can be written as 
\begin{equation} 
H_{1\, band}^{(1)} = U \sum_{i,j,\sigma} \hat {v}^{}_{ij\sigma} 
(\alpha_{i\bar{\sigma}}^\dagger \beta^{}_{j\bar{\sigma}}+  
\beta_{j\bar{\sigma}}^\dagger  \alpha^{}_{i\bar{\sigma}}) 
\label{lot} 
\end{equation} 
with 
\begin{equation} 
\hat {v}_{ij\sigma}=\sum_{l,l',n} b^{}_{jn} a^{}_{in} a^{}_{ln} 
a^{}_{l'n} \alpha_{l\sigma}^\dagger  
\alpha^{}_{l'\sigma} 
\label{v} 
\end{equation}	 
Here the terms with $l\neq l'$ can be neglected when the 
lower band is flat so we can rewrite Eq.~\ref{v} as 
\begin{equation} 
\hat {v}_{ij\sigma} \sim -n_{i\sigma}^\alpha\frac{V}{|\epsilon_f|} ( \delta_{i,j} 
+\frac{t}{|\epsilon_f|} \delta_{|i-j|,1}) 
\label{vijs} 
\end{equation} 
Thus the lowest order inter-band process are correlated hoppings 
between $\alpha_i$ and $\beta_j$ orbitals and are proportional to the 
spin polarization (opposite to the spin of the electron which hops) 
surrounding the $j$ site: the hoping occurs out of regions of 
ferromagnetically aligned electron spins. 
 
It is interesting to note that the origin of this ferromagnetic 
alignment is not an exchange mechanism but processes involving 
charge-transfer. To see this more clearly we show in the Appendix that 
by using a Schrieffer-Wolff transformation we can reduce the 
multi-band term to an effective one band term 
 
\begin{eqnarray} 
\hat {H}^{(2)}_{1\, band}&=&\frac{1}{2}[\hat {T}_1,H^{(1)}_{1\, band}]  
\nonumber\\  
&=& -4\frac {U^2V^2} {N|\epsilon_f|} \sum_{i,j, \bold{k},\sigma}  
\frac {n^{\alpha}_{i\sigma} n^{\alpha}_{j\sigma}  
(\alpha_{i\bar{\sigma}}^\dagger \alpha^{}_{j\bar{\sigma}}+\alpha_{j \bar{\sigma}}^\dagger \alpha^{}_{i\bar{\sigma}})} 
{E^{+}_{\bold{k}}- \tilde \epsilon_f-U} \tilde {t}_{ij}(\bold{k}) 
\end{eqnarray} 
where  
\begin{eqnarray} 
\tilde {t}_{ij}(\bold{k}) = \left\{\begin{array}{ll} 
1, & {\rm for}\ i=j \\ 
\frac {t}{|\epsilon_f|} s_{\bold k}, & \mbox{for  
$i$ and $j$ n. n.} \\ 
\frac {t^2}{|\epsilon_f|^2} s_{\bold k}^2,  &\mbox{for $i$ and $j$ second and third n. n.} 
\end{array} 
\right.\nonumber\\ 
\end{eqnarray} 
with $s_{\bold k}=\cos{k_x}+\cos{k_y}$. The hopping between two $\alpha$ 
orbitals is only possible if both sites are occupied with electrons 
having the same polarization.  Then it is clear that 
${H}^{(2)}_{1\, band}$ induces a ferromagnetic interaction between the 
localized $f$-state which comes from the itinerancy of the carriers trough 
the upper band. In addition, to maximize the energy gain, the added 
carriers must occupy ${\bf k}\sim 0$ states. In this way, the 
magnitude of the hopping ${t}_{ij}(\bold{k})$ is maximized at the same 
time the denominator $E^{+}({\bf k})-\tilde {\epsilon_f}-U$ is 
minimized (tends to $\Delta-U$).  The electrons can be added to  ${\bf 
k}\sim 0$ states only if the background is ferromagnetic and, of 
course, opposite to the spin of the added electron. These 
charge-transfer processes enhancing ferromagnetism involve the states 
in the lower part of the upper band. 
 
\begin{figure}[tbp] 
\begin{center}
\epsfig{file=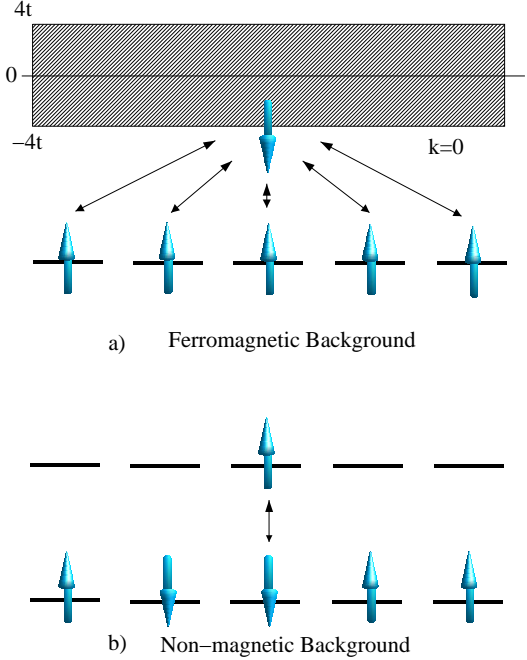,width=70mm,angle=0} 
\end{center}
\caption{Conduction electron compensation of $f$-orbital moments. (a) 
The mechanism for the effective one-band Hamiltonian. An electron at 
the bottom of the conduction band is partially compensating the 
collecive ferromagntic state. This process more effectively lowers 
conduction electron kinetic energy than mechanism (b). In the latter 
the conpensation is a one-to-one on-site process and is present for an 
anti-ferromagneitc or paramagnetic aligment of the $f$-orbitals. A 
paramagnetic alignment is shown} 
\label{fig2} 
\end{figure}

This inter-band process is illustrated in Fig.~\ref{fig2}a. The 
representation emphasizes the collective nature of the mechanism. The 
virtually hopping electron has reached a ${\bf k}=0$ band state. It is 
anti-aligned with the $f$-orbitals which are ferromagnetically aligned 
among themselves. The moment of this band state partially compensates 
the fully saturated ferromagnetic alignment of these orbitals. This 
compensation picture differs from ``the exhaustion picture''of 
Nozi\'eres \cite{Nozieres,Nozieres2} evoked by Tahvildar-Zadeh {\it 
et al.\/} \cite{Thavildar-Zadeh} The compensation is on a collective 
state to collective state basis and not the collective state to single 
moment basis argued by Nozi\'eres. This difference highlights the 
difficulty building the physics of the periodic Anderson model from 
the physics of its single impurity version.  
 
The process in Fig.~\ref{fig2}b contrasts that in 
Fig.~\ref{fig2}a. This process in Fig.~\ref{fig2}b compensates on 
a site-to-site basis and is the one present for a paramagnetic or an  
anti-ferromagnetic alignment of the 
$f$-orbitals. The energy cost for virtual hopping is higher in than 
that in Fig.~\ref{fig2}a. This leads to larger energy denominators in 
perturbation theory and in turn leads to a smaller lowering of the 
ground state energy. 
 
\subsection{Effective Hamiltonian: Canonical Transformation} 
 
Here we will present an effective low energy model valid in a 
different region of parameter space: $U\gg W/2,-\epsilon_f\gg V>0$. 
In this regime of parameters and around 1/4 filling, single electron 
occupancy of the $f$-states occurs mainly because of the double 
occupancy penalty of the Coulomb repulsion. As we will see, the 
interaction between the moments in these $f$-states is dominated by the 
RKKY interaction. 
 
To derive this effective Hamiltonian, we will make a fourth order 
Schrieffer-Wolff transformation \cite{Schrieffer} as in 
Ref.~\cite{Zhou}: 
\begin{equation} 
\tilde {H}= e^{\hat{W}_3} e^{\hat{W}_1} H e^{-\hat{W}_1} e^{-\hat{W}_3} 
\label{Hmon} 
\end{equation} 
where the transformation operators $W_1$ and $W_3$ are of order $V$ 
and $V^3$. With this transformation we get a new Hamiltonian 
$\tilde{H}$ without terms of order $V$ and $V^3$.  By means of another 
canonical transformation we eliminate the term of order $V^2$. In this 
way we get the low energy effective Hamiltonian $H_{spin}$ correct 
through $V^4$. The details of the derivation are given in the 
Appendix. 
 
The final expression for the effective Hamiltonian is a Heisenberg 
Hamiltonian 
\begin{equation} 
H_{spin}= \sum_{ij} J_{ij} {{\bf S}_i\cdot {\bf S}_j} 
\end{equation} 
where  
\begin{equation} 
J_{ij}=J_{ij}^{(0)}+J_{ij}^{(1)}+J_{ij}^{(2)}+ J_{ij}^{(3)}+ 
J_{ij}^{(4)} 
\label{eq:exchange} 
\end{equation} 
The RKKY contribution $J_{ij}^{(0)}$ is given by 
\begin{equation} 
J_{ij}^{(0)} = {\frac{V^4}{4N^2}}\sum_{\bf k,k'} e^{-i 
({\bf k-k}')\cdot {\bf R}_{ij}} 
\frac {2\langle n^d_{\bold k}\rangle_0(1-\langle n^d_{\bold k'}\rangle_0)} {e_{\bf k}-e_{\bf k'}} (\gamma_{\bf 
k}+\gamma_{\bf k'})^2 
\label{RKKY} 
\end{equation} 
where 
\begin{equation} 
\gamma_{\bf k} = \frac{1}{e_{\bf k}-\epsilon_f-U}-\frac{1}{e_{\bf 
k}-\epsilon_f} 
\end{equation} 
$\langle n^d_{\bold k}\rangle_0$ is defined in the Appendix.  
 
$J_{ij}^{(1)}$ is associated with virtual processes where one 
$f$-electron at sites $i$ and $j$ go through the conduction band, 
doubly occupy the $d$-sites $j$ and $i$, and one of these two 
electrons comes back to an empty $f$-orbital. The expression for 
$J_{ij}^{(1)}$ is 
\begin{equation} 
J_{ij}^{(1)}= {\frac{V^4}{UN^2}}\sum_{\bold {k},{\bf k'}} e^{-i 
({\bf k-k}')\cdot {\bf R}_{ij}} (2 \delta_{\bold k}+\gamma_{\bold k})(2 
\delta_{\bold k'}+\gamma_{\bold k'}) 
\end{equation} 
The other contributions to $J_{ij}$ are given by 
\begin{eqnarray} 
J_{ij}^{(2)}&=&{\frac{V^4}{4N^2}}\sum_{{\bf k}, {\bf k'}}  
                e^{-i({\bf k-k'})\cdot {\bf R}_{ij}} 
            (3 \delta_{\bold k}\delta_{\bold k'} m_{\bold kk'} 
               + x_{\bold kk'} y_{\bold {kk'}})\\ 
J_{ij}^{(3)}&=&{\frac{V^4}{4N^2}}\sum_{\bold {k},{\bf k'}} 
                e^{-i({\bf k-k'})\cdot {\bf R}_{ij}} 
     [ \gamma_{\bold k}\gamma_{\bold k'} n_{\bold kk'}\nonumber\\ 
     & & \hspace{1.5cm} + 4 m_{\bold kk'} y_{\bold {kk'}} 
         - 7 \gamma_{\bold k}^2 \gamma_{\bold k'}  
          (3 \gamma_{\bold k}+\gamma_{\bold k'})]\\ 
J_{ij}^{(4)} &=& {\frac{V^4}{2N^2}}\sum_{\bold {kk'}}  
                e^{-i({\bf k-k'})\cdot {\bf R}_{ij}}  
                \gamma_{\bold k} \gamma_{\bold k'} m_{\bold kk'}^2 
\end{eqnarray} 
where $m_{\bold {kk'}}$, $n_{\bold {kk'}}$, $x_{\bold {kk'}}$ and 
$y_{\bold {kk'}}$ are defined in the Appendix. These four contributions to 
$J_{ij}$ are about one order of magnitude smaller than 
$J_{ij}^{(0)}$.  
 
We expressed the magnitude of the effective spin-spin interactions 
$J_{ij}$ (\ref{eq:exchange}) in terms of ${\bf k}$-space states because in 
the absence of hybridization they are the eigenstates of the 
problem. From this point of view, the ferromagnetism for low density 
of carriers comes from the RKKY term $J^{(0)}_{ij}$ and the small 
volume enclosed by the Fermi surface: When the Fermi volume is small, 
the transferred ${\bf k}$-vector between different points on the surface 
is small, and the relevant phase factors of Eq.~\ref{RKKY} are 
positive for small distances between sites $i$ and $j$. This 
positivity gives an effective ferromagnetic interaction between near 
neighbors, which is proportional to the density of carriers and 
competes with the usual anti-ferromagnetic super-exchange interaction 
included in the other components of $J_{ij}$. Because the 
ferromagnetic interaction increases with the concentration of 
carriers, we expect the appearance of a ferromagnetic phase above some 
critical concentration ${\rho_c}$. Due to the non-interacting case 
($U=0$) being paramagnetic, we also expect a ferromagnetic phase above 
some critical value $U_c$. To derive a simple expression for the phase 
boundary $U_{c} \equiv U({\rho_{c}})$, we have to compare the 
effective ferromagnetic interaction $J^{(0)}_{ij}$ with the the 
antiferromagnetic super-exchange interaction, which is proportional to 
$V^{4}t^{2}/U^{5}$ for the symmetric PAM.  From 
Eq.~\ref{RKKY} it is clear that, for the dilute case, $J^{(0)}_{ij}$ 
is proportional to the concentration of carriers $\rho_c$ and 
$V^4/tU^2$.  Consequently, the phase boundary of the ferromagnetic 
region is given by 
\begin{equation} 
{\rho_c} V^4/tU^2 \sim V^4t^2/U^5  
\end{equation} 
which implies that 
\begin{equation} 
{\rho_c}\sim t^{3}/U_{c}^{3} 
\end{equation} 
This expression qualitatively agrees with the numerical results of 
Ref.\cite{Guerrero2}.   
 
This ferromagnetic phase must disappear above some upper critical 
concentration of carriers where the volume enclosed by the Fermi 
surface is no longer small. For this concentration we do not expect an 
ordered phase unless nesting is present. In presence of nesting, 
$E^-({\bf k})=E^-({\bf k}+{\bf Q})=E_{\rm Fermi}$, the Fermi surface 
may be unstable towards the development of a spin density wave. This 
instability is manifested in Eq. ~\ref{RKKY} from the divergence of 
the ${\bf Q}$ Fourier component of $J^{(0)}_{ij}$.  For a commensurate 
state, ${\bf Q}$ is a high symmetry point of the Brillouin zone and 
equals one-half of a reciprocal lattice vector. For a square lattice 
two such points exist: $(0,\pi)$ for $\rho=1/4$ and $(\pi,\pi)$ for 
$\rho=1/2$.  In the absence of nesting no well defined ${\bf Q}$ 
vector connects different points of the Fermi surface, and preferred 
ordering for the spins is absent.  For this reason, if we were to 
increase $\rho$ over the upper ferromagnetic phase boundary, we would 
expect a paramagnetic phase for any concentration different from 1/4 
and 1/2.  This picture is similar to the one already obtained in 1D 
\cite{Guerrero}.  The only difference between 1D and 2D is in 1D 
nesting exists at the Fermi level for any concentration of 
carriers. This nesting explains why spiral critical order at $2{\bf 
k}_F$ is obtained for any value of $\rho$ in 1D\cite{Guerrero}.

\subsection{Quantum Monte Carlo Results} 
 
All our simulations were performed for lattices of $4\times 4$ and 
$6\times 6$ unit cells. The cost of performing these simulations is 
approximately the same as simulating an $8\times 8$ and $12\times 12$ 
one-band Hubbard model.  As mentioned previously, we used ground 
states derived from the non-interacting problem as 
$|\psi_T\rangle$. When we simulated the PAM with $\epsilon_f=-5$, we 
used the non-interacting states for $\epsilon_f=-2$ because this choice 
consistently produced a lower estimate of the ground state energy with 
a smaller statistical error then we would obtain if we had used the 
non-interacting states for $\epsilon=-5$. 
 
Most of our simulations were performed for electron fillings of 1/4 
through 3/8. In our previous work \cite{Bonca}, where we studied the 
region from 3/8 to 1/2 and found the RSDW ground state at 3/8 filling 
for a $6\times 6$ system. At 1/2 filling, we found the expected AF 
ground for $U\gtrsim 2$ in agreement with other QMC simulations. In 
between, we found a paramagnetic (PM) ground state. For these previous studies, 
we had $\epsilon_f=-2$. As part of the present study, we repeated some 
simulations over this 3/8 to 1/2 range with $\epsilon_f=-5$ but did 
not find any indications of a FM state. We note again that the 
predictions of DMFT and SBMFT are inconsistent with these PM and RSDW 
states. The effective Hamiltonians presented in this paper are 
expected to be inappropriate for this filling range. We will not 
discuss this range further. 
 
In the present work, we also did a series of rough simulations at 
fillings less than 1/4 for various values of $U$ and $\epsilon_f$. We 
only found PM ground states. DMFT and SBMFT find FM ground states for 
identical parameter choices.

\begin{figure}[tbp] 
\begin{center}
\epsfig{file=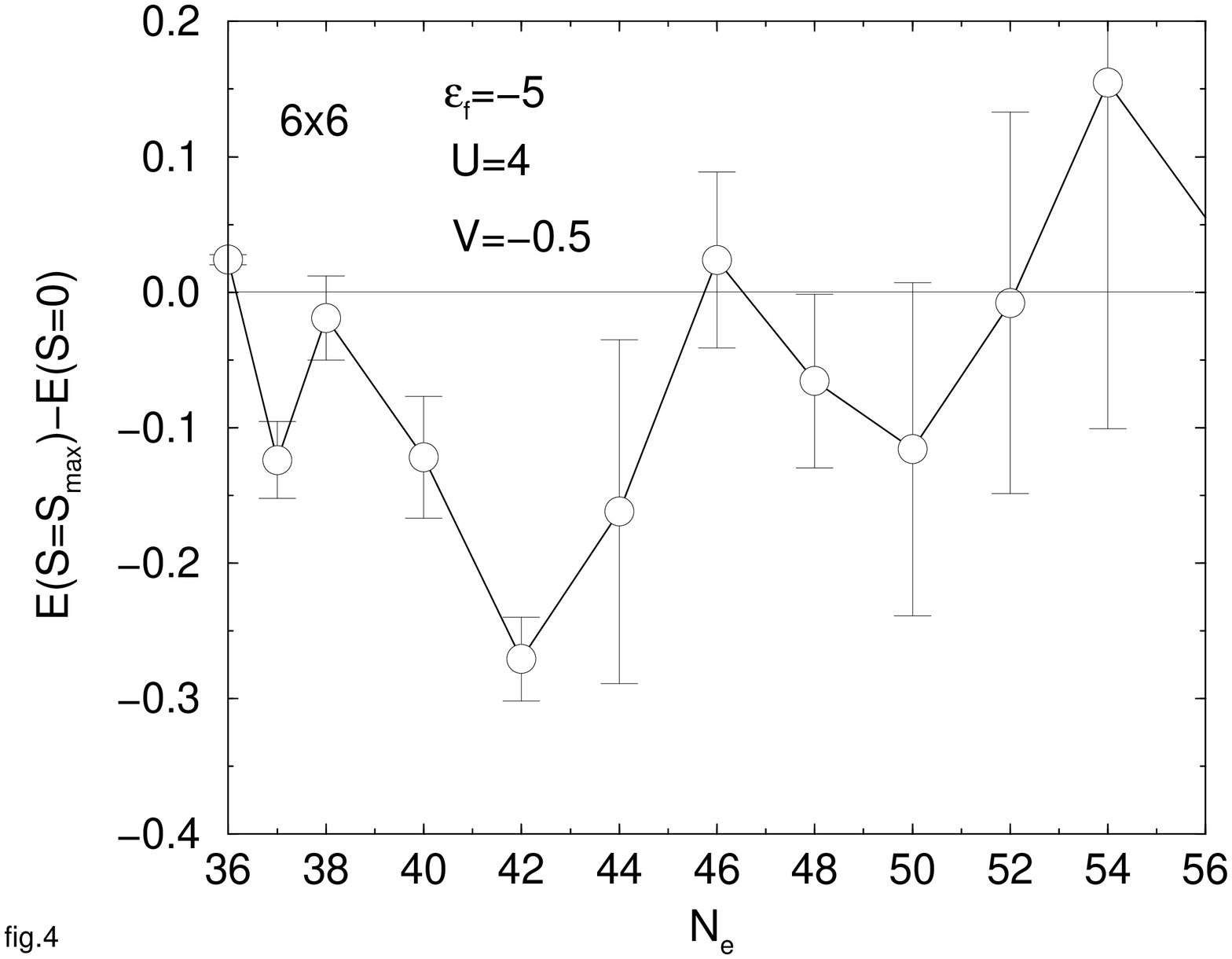,width=65mm,angle=-90} 
\end{center}
\caption{Energy difference between the partially polarized and singlet  
ground state energies as a function of electron filling. The lattice 
has $6\times 6$ unit cell and the model has $U=4$, $V=-0.5$, and 
$\epsilon_f=-5$.} 
\label{fig4} 
\end{figure}

We note that we found a PM ground state for one electron removed from 
1/4 filling, even for large values of $U$. As we will discuss below, 
when one electron was added to 1/4 filling, we found a FM state for 
$\epsilon_f=-5$. Clearly the properties of the PAM are asymmetric 
about 1/4 filling in contrast to generally accepted suggestions that 
it behaves as a half-filled, nearest neighbor hopping Hubbard model 
which displays particle-hole symmetry.

\begin{figure}[tbp] 
\begin{center}
\epsfig{file=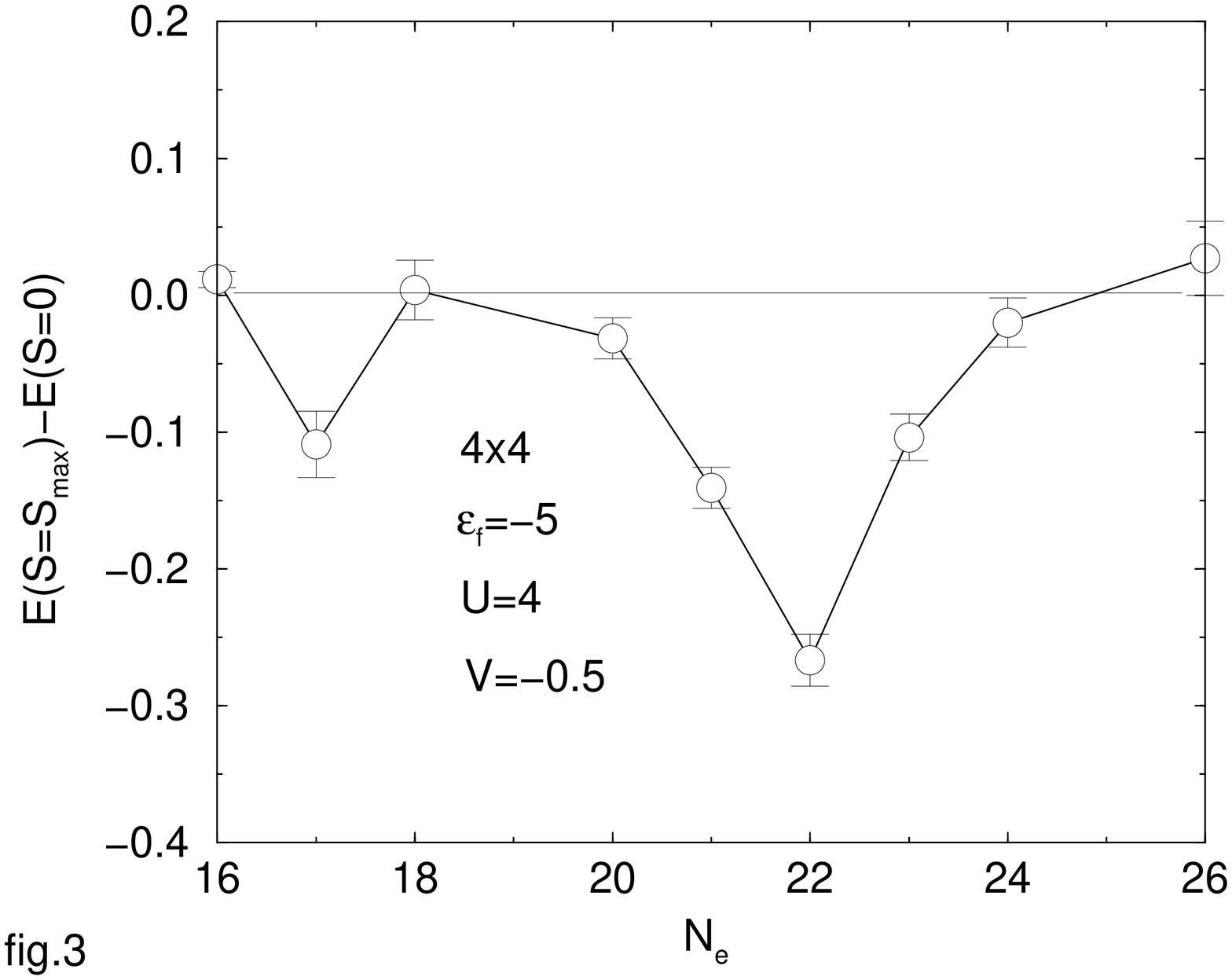,width=65mm,angle=-90} 
\end{center}
\caption{Energy difference between the partially polarized and singlet  
ground state energies as a function of electron filling. The lattice 
has $4\times 4$ unit cell and the model has $U=4$, $V=-0.5$, and 
$\epsilon_f=-5$.} 
\label{fig3} 
\end{figure}

Figures~\ref{fig3} and \ref{fig4} show the main results of our 
simulations. They plot the energy difference between a polarized and a 
singlet ground state as a function of electron 
filling. Figure~\ref{fig3} is for the $4\times 4$ lattice, and 
Fig.~\ref{fig3} is for the $6\times 6$ lattice. In both figures, 
$t=1$, $V=0.5$, $U=4$, and $\epsilon=-5$. For the non-interacting 
problem, these parameters are the ``flat band'' case illustrated in 
Fig.~\ref{fig1}b.

At 1/4 filling, the two energies are equal to within statistical 
error, indicating the ground state is likely PM. This is not what is 
expected. Our effective one-band Hamiltonian has long range 
hopping. So for a sufficiently large value of $U$, we expect an AF 
ground-state as was observed in the one-dimensional DMRG 
calculations. Calculating $J_{\rm RKKY}$ for the effective Heisenberg 
model, we found that when $i\ne j$ the magnitude of the exchange 
interaction is about two orders of magnitude smaller than the 
magnitude of our statistical error. Thus we should not expect to see 
this state easily. If we used an anti-ferromagnetic state for 
$|\psi_T\rangle$, we would see a AF ground state, but the energy of 
this biased result and the one from a PM $|\psi_T\rangle$ were 
typically equal within statistical error. 

By doping 1/4 filling with one extra election, we found that the FM 
state has a lower energy. Adding one electron more produced a PM 
state. This behavior with the doping of one and two electrons is 
reminiscent of Nagaoka ferromagnetism and its instability in the 
$U=\infty$ nearest-neighbor-hopping Hubbard model. While our model is 
not the same, the proposed mechanism for ferromagnetism is 
very similar: An electron lowers its kinetic energy by moving 
through a ferromagnetic background. In our mechanism, the electron lowers 
its kinetic energy by inter-band processes enabling its hopping 
between two occupied $f$-states anti-aligned with its spin. The need 
for inter-band processes differentiates our mechanism from 
Nagaoka's. 
 
\begin{figure}[tbp] 
\begin{center}
\epsfig{file=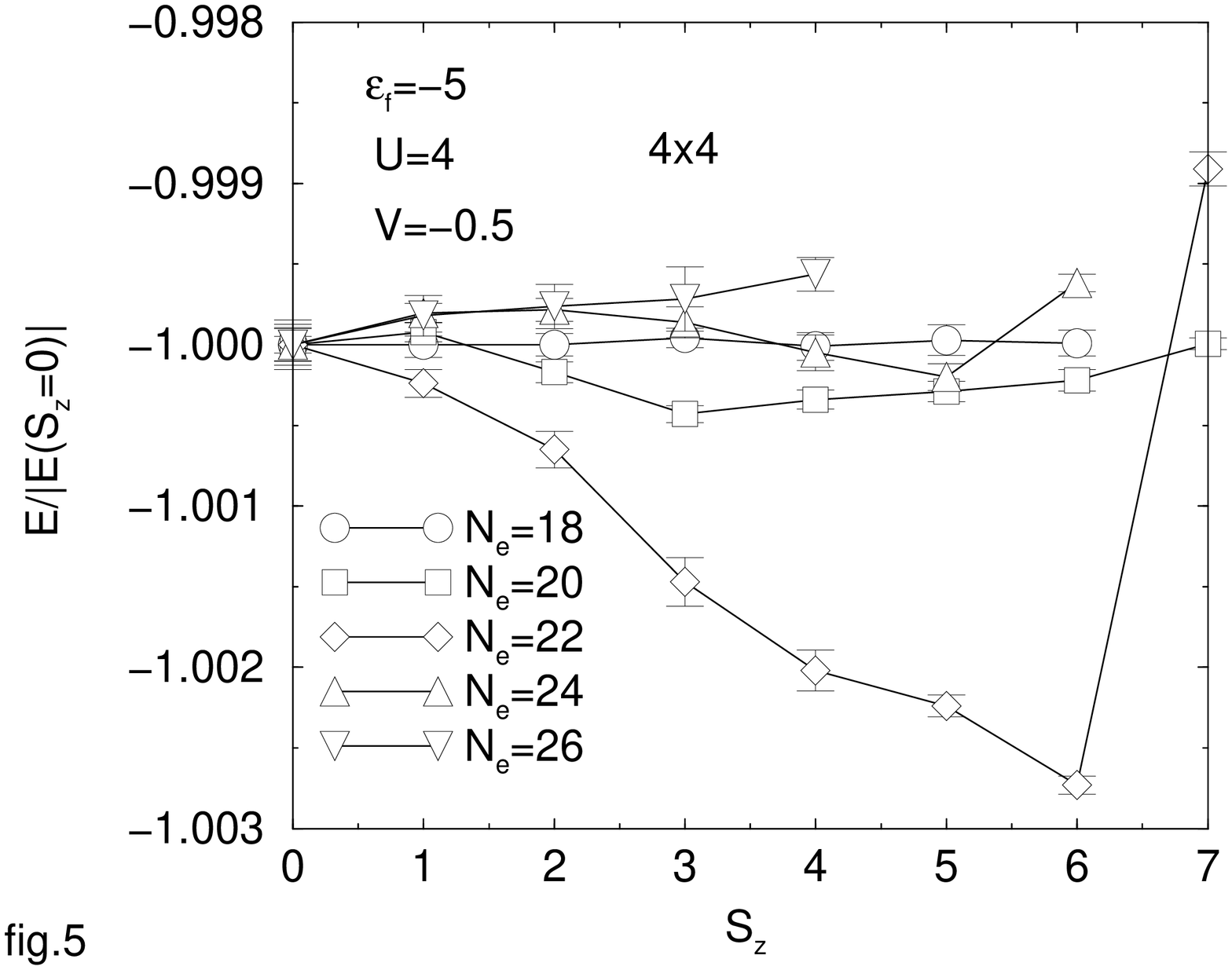,width=65mm,angle=-90} 
\end{center}
\caption{Ground state energy as a function of the value of the total spin 
$S_z$ for several different values of electron filling. The lattice 
has $4\times 4$ unit cell and the model has $U=4$, $V=-0.5$, and 
$\epsilon_f=-5$.The number of electrons varies as 18,  
20, 22, 24, and 26.} 
\label{fig5} 
\end{figure}

We believe the disappearance of the FM with the second electron is not 
analogous to the known instability of the Nagaoka state but rather is 
a finite-size effect. It is well documented that the shell structure 
of a finite-sized non-interacting problem is reflected in the behavior 
of the filling dependence of the energy in the interacting problem 
\cite{Furukawa,Guerrero3,Guerrero4}. 
At least to a first approximation, the chemical potential is constant 
in a shell and is discontinuous between such shells. According to 
our effective one-band model, if we were to add one electron to the 
1/4-filled case, it would most effectively lower its kinetic energy by 
moving between oppositely aligned $f$-states by virtually hopping 
through a ${\bf k}=0$ conduction (upper) band state. The Pauli 
principle blocks, or frustrates, this process for a second electron of 
the same spin if finite-size effects make the ${\bf 
k}=(\pm\pi/2,\pm\pi/2)$ states of the next shell energetically 
inappropriate: The second electron would have to enter the ${\bf k}=0$ 
shell oppositely aligned with the first, and the ferromagnetic 
background of $f$-states cannot accommodate the two different 
alignments. 
 
The influence of finite-size and inter-band process is also seen in 
Fig.~\ref{fig3} where we show the energy ratio $E/|E(S_z=0)|$ as the 
number of electrons increases from 18 to 26 electrons, from one 
closed conduction band shell, $N_e=18$, through another, $N_e=24$, and 
ending at an open conduction electron shell. At first the polarized 
ground state becomes much lower than the PM state and then two become 
approximately equal at $N_e=24$. This variation is highlighted in 
Figs.~\ref{fig5}a-e which show the ground state energy as a function 
of $S_z$ for different electron fillings. For 18 and 26 electrons, the 
energy per site either does not vary within statistical error or 
increases. For 20, 22, and 24 electrons, it has a clear minimum but 
the minimum for 24 electrons is just barely below unity. We note that 
the minimum in energy and the maximum in $S_z$ occurs for a 
half-filled conduction electron shell. It is as if a Hund's rule 
coupling is operative in a non-multi-orbital situation. In 
Fig.~\ref{fig3}, $S_{max}$ is the value of $S_z$ for which 
$E/|E(S_z=0)|$ is a minimum. If this ratio had no clear minimum, then 
we used $S_{max}=\frac{1}{2}(N_e-N_s)$ where $N_s$ is the number of 
lattice sites (the number of $f$-orbitals) and $N_e$ is the number of 
electrons. 
 
$N_e=24$ corresponds to 3/8 filling. Instead of being a possibly a 
weak FM state, we argue that it is actually an unpolarized state RSDW 
state; however, while we do not clearly see this state for the 
$4\times 4$ lattice, we clearly saw it for the $6\times 6$ lattice 
when $N_e=56$ \cite{Bonca}. There is an important and subtle difference between the 
two lattices: Our performance gain in constructing $|\psi_T\rangle$ is 
based on using closed shell states of the {\em non-interacting 
problem\/}, and these states mix information about both the valent and 
conduction bands. For these states, the $N_e=24$ singlet for $4\times 
4$ is a half-filled shell, but the $N_e=56$ singlet for $6\times 6$ is 
a closed shell. This difference makes it much more difficult to see 
the RSDW state for the $4\times 4$ lattice. We emphasize that the 
shell in Fig.~\ref{fig3} is different. It is a remnant of a shell in 
the {\em conduction} band.

We remark that our effective Heisenberg models admits a RSDW state. In 
particular, within the effective Heisenberg model, the ferromagnetic 
nature of the RKKY exchange interaction depends on the volume enclosed 
by the Fermi surface being small. Increasing the number of electrons 
increases the volume, lengthens the {\bf k}-vector for transfer across 
the Fermi surface, and decreases the strength of the interaction until 
the Fermi surface becomes unstable because of nesting. Direct 
calculation of the Fourier components of $J_{ij}^{(0)}$ shows that it 
diverges close to nesting. The spatial pattern of this diverging 
Fourier component corresponds the $J_{ij}^{(0)}$ being zero if sites 
$i$ and $j$ are on different sublattices and behaving 
anti-ferromagnetically if the sites are on the same sublattice. This 
is precisely the pattern observed in the spin-spin correlation 
function reported in our previous work, but we do not see it for the 
$4\times 4$ lattice size. Finite sizes effects and the limited range 
of system sizes we can afford to simulate have prevented us from 
performing the scaling analysis need to establish the RSDW state as 
one of long-range order. 
 
Both the ferromagnetism and conduction band shell structure is evident 
for the $6\times 6$ case shown in Fig.~\ref{fig4}. Adding one electron 
to the 1/4 filled case ($N_e=36$) produces a FM state. Adding another 
closes the ${\bf k}=(0,0)$ shell which frustrates the 
ferromagnetism. Adding more electrons to reach $N_e=54$ successively 
populates the ${\bf k}=(\pm\pi/3,\pm\pi/3)$ and $(\pm 2\pi/3,\pm 
2\pi/3)$ conduction band shells. At the middle of each shell, the 
energy difference is a maximum and the shell is half filled.

\section{Conclusions} 
 
Our numerical results indicate that several important features of the 
phase diagram of the one dimensional PAM\cite{Guerrero2,Guerrero} are 
preserved in two dimensions. In both dimensions and for half-filling, the 
Coulomb interaction induces an insulating gap, and the system can have 
AF order (AF insulator), or it can remain in a paramagnetic state 
(Kondo insulator) if there is a strong enough hybridization $V$ 
between both bands. In this latter case there is also a spin gap 
associated with the energy necessary to break a Kondo singlet. The AF 
order originates in the Fermi surface nesting at ${\bf Q}=(\pi,\pi)$. 
 
When doped away from half-filling, the system in two dimensions 
becomes paramagnetic.  In one dimension, however, critical 
inconmensurate correlations peaked at ${\bf Q}=2{\bf k}_f$ develop. 
This can be understood with the effective Heisenberg theory derived in 
Section IV.B: The RKKY interaction $J^{(0)}_{ij}$ has a divergent 
Fourier component at ${\bf Q}=2{\bf k}_f$ because there is nesting for 
any concentration of electrons. This results in a Luttinger 
liquid with spin-spin correlations which are critical and peaked at 
${\bf Q}=2{\bf k}_f$.  The situation is different in two dimensions 
where there is nesting only for fillings $1/2$, $3/8$, and close to 
1/4 (small Fermi surface). Our numerical evidence suggests that the 
system is paramagnetic between $3/8$ and $1/2$ filling. For $3/8$ 
filling, where the nesting appears at ${\bf Q}=(0,\pi)$ and $(\pi,0)$, 
our CPMC results indicate the presence of a RSDW phase.  In this 
phase, the two interpenetrating sublattices are decoupled, and there 
is AF order in each. This spatial ordering can be also understood by 
considering the effective Heisenberg theory for the PAM: It is clear 
that in going from $1/2$ to $1/4$ filling the nearest neighbor RKKY 
interaction changes its sign (AF close to $1/2$ and FM close to 
$1/4$). A cancellation then must occur at some intermediate 
concentration. We can easily see that this intermediate filling is 
$3/8$ where there is nesting for two different wave vectors: ${\bf 
Q}=(0,\pi)$ and $(\pi,0)$.  These two transferred wave vectors 
give canceling contributions to the nearest neighbor RKKY 
interactions. For instance, the nearest neigbhor in the $x$-direction 
will feel an antiferromagnetic interaction coming from ${\bf 
Q}=(\pi,0)$ and a ferromagnetic one of the same magnitude coming from 
${\bf Q}=(0,\pi)$. On the other hand, the next nearest neighbors have 
an overall phase $\pi$ for both wave vectors (constructive 
interference) and the effective interaction is therefore AF. This is 
the origin of the RSDW state obtained for $3/8$ filling. 
 
By decreasing the filling below $3/8$, the CPMC results indicate that 
the system becomes paramagnetic again down to some critical filling 
beyond which non-saturated ferromagnetism appears. Again this behavior 
is related to the nesting of the Fermi surface. Below $3/8$ filling 
there is no nesting down to some small concentration of conduction 
electrons near 1/4 filling where the Fermi surface of the conduction 
electrons can be very well approximated by a small sphere and the wave 
vector $2{\bf k}_f$ connecting two different points is very close to 
zero. Under these conditions, the RKKY interaction $J^{(0)}_{ij}$ 
diverges with a negative value indicating the presence of a 
ferromagnetic instabilty.   
 
This mechanism is not the only one giving rise ferromagnetism in the 
PAM for small concentrations of carriers. In the Mott insulating 
regime ($U<\Delta$) where the one band effective model $H_{1\, band}$ 
is a valid effective low energy theory, the ferromagnetism is related 
to the long range hopping processes involving charge virtually 
tranferring from the lower to the upper band. The relevance of these 
processes is related to the flatness of the lower band and the 
dispersive character of the upper one. Without considering processes 
connecting the two different bands ($H_{1\, band}=H^{(0)}_{1\, 
band}$), the effective one band Hamiltonian is a an extended Hubbard 
model (the hopping is not restricted to nearest neighbors). Due to the 
flatness of the lower band, the effective ratio ${\tilde 
U}/{\tau^{\alpha}_{ij}}$ is very large, and the model is in a 
Nagaoka-like region even for small values of the bare interaction 
$U$. The relevant low energy scale which determines the magnetic 
ordering comes from the comparison between $\rho \tau^{\alpha}_{ij}$ 
(energy per link of the Nagaoka state) and ${\tau^{\alpha}_{ij}}^2/ 
{\tilde U}$ (energy per link of the AF state).  As this difference is 
extremely small, any other term added to the Hamiltonian, which favors 
one of the two competitive phases, will be relevant. As we have 
explained in Section IV.A, $H^{(2)}_{1\, band}$ stabilizes the 
ferromagnetic phase.  If the localized electrons are polarized in the 
same direction, the added carriers can gain energy from virtual 
processes transferring charge between ${\bf k}\sim 0$ states of the 
lower and the upper band. 
 
It is also important to make some additional comments about the finite 
size effects in our numerical results.  In an infinite system it is 
necessary to have a finite critical concentration of electrons in 
order to induce the ferromagnetic phase.  We argue that This must be 
the case because the gain in kinetic energy of the added particles 
must overcompensate the loss of magnetic energy of the localized 
electrons (which is proportional to the system size). The numerical 
results show that the system becomes ferromagnetic with the addition 
of only one conduction electron. This is because one electron added to 
a $4\times4$ or $6\times6$ system corresponds to a finite 
concentration.  Besides, as it is explicitly shown in Figs.~\ref{fig3} 
and \ref{fig4}, there are closed shell effects that give a 
non-monotonic behavior for the energy difference between the 
ferromagnetic and the paramagnetic state as function of 
concentration. It is clear that for small systems and closed shell 
conditions, the non-interacting (paramagnetic) solution will be more 
stable under the introduction of correlations. The small systems will 
generally have larger energy gaps between two shells.  
 
Due to size effects, we must assume that we are dealing with states of 
long-range order, and we cannot say much about the order of the 
paramagnetic-ferromagnetic transitions. To estrablish long-range order 
and to determine the order of the transitions, it is necessary to 
scale the magnetization with the system size for a fixed filling. To 
do this properly requires larger systems than the ones considered in 
this paper.

%
 
%
%
 
\acknowledgments 
 
We thank M. Guerrero, M. Gul\'asci, M. Jarrell, Th. Pruschke, and 
G. Ortiz for useful discussions. Work at Los Alamos is sponsored by 
the US DOE under contract W-7405-ENG-36. 
 
\appendix  
\section*{A} 
 
To make the derivation of $H_{spin}$ clearer, we begin by 
rewriting Eq.~\ref{eq:pam} as 
\begin{equation} 
H=H^{(0)}+H^{(1)} 
\end{equation} 
with  
\begin{equation} 
H^{(0)}= \sum_{\bold {k},\sigma} e^{}_{\bold k} 
d_{\bold{k}\sigma}^\dagger d^{}_{\bold{k}\sigma} 
+\epsilon^{}_f\sum_{i,\sigma}n_{i\sigma}^f+\frac{U}{2} 
\sum_{i,\sigma}n_{i\sigma}^fn_{i\bar {\sigma}}^f 
\end{equation} 
\begin{equation} 
H^{(1)}=V\sum_{i,\bold{k},\sigma} (e^{i{\bf k}\cdot {\bf r}_i}d_{{\bold 
k}\sigma}^\dagger f^{}_{i\sigma} +e^{-i 
{\bf k}\cdot {\bf r}_i}f_{i\sigma}^\dagger d^{}_{\bold{k}\sigma}) 
\end{equation} 
 
The next step is the application to $H$ of a Schrieffer-Wolff-like 
transformation \cite{Schrieffer} to eliminate terms of order $V$ and 
$V^3$ from the Hamiltonian.  We do this by expanding the exponentials 
of Eq.~\ref{Hmon} and imposing the conditions 
\begin{equation} 
H^{(1)}+[\hat W_1,H^{(0)}]=0 
\label{cond1} 
\end{equation} 
\begin{equation} 
\frac {1}{3} [\hat {W}_1,[\hat {W}_1,H^{(1)}]]+[\hat W_3,H^{(0)}]=0 
\label{cond2} 
\end{equation} 
that define $\hat{W}_1$ and $\hat{W}_3$. With these conditions 
\begin{equation} 
\tilde {H}=H^{(0)}+\frac{1}{2}[\hat {W}_1,H^{(1)}] 
   +\frac{1}{8}[\hat{W}_1,[\hat {W}_1,[\hat {W}_1,H^{(1)}]]]+\cdots  
\end{equation} 
 
The following expression for $\hat {W}_1$ satisfies Eq.~\ref{cond1} 
\begin{equation} 
\hat {W}_1=V \sum_{i, \bold{k},\sigma}  
 (\delta^{}_{\bold k}+\gamma^{}_{\bf k} n^{}_{i\bar \sigma})  
 (e^{i{\bf k}\cdot {\bf r}_i}d_{{\bold k}\sigma}^\dagger f^{}_{i\sigma} 
 -e^{-i{\bf k}\cdot {\bf r}_i}f_{i\sigma}^\dagger d^{}_{\bold{k}\sigma}) 
\end{equation} 
with 
\begin{equation} 
    \gamma_{\bf k}= 
    \frac{1}{e_{\bf k}-\epsilon_f-U}-\frac{1}{e_{\bf k}-\epsilon_f} 
\end{equation} 
and 
\begin{equation} 
   \delta_{\bf k}= 
       \frac{1}{e_{\bf k}-\epsilon_f} 
\end{equation} 
With this $\hat{W}_1$, the second order term in $\tilde H$ becomes 
\begin{eqnarray} 
H^{(2)}&=&\frac{1}{2}[\hat {W}_1,H^{(1)}]\nonumber\\ 
&=& - \frac{V^2}{2} \sum_{i,j,\sigma} e^{i({\bf k-k'})\cdot {\bf r}_i} 
[t_{ij}^{(1)}+t_{ij}^{(2)}(n^{}_{i\bar\sigma}+n^{}_{j\bar\sigma})] 
f_{j\sigma}^{\dagger} f^{}_{i\sigma} 
\nonumber\\ 
&-& V^2 \sum_{i,\bold{k},{\bf k'},\sigma, \sigma'} e^{i 
({\bf k-k}')\cdot {\bf r}_i} m_{\bf kk'} 
{\bf S}^{}_i \cdot d_{{\bf k}\sigma}^{\dagger} {\bf s}^{}_{\sigma 
\sigma'}  d^{}_{{\bf k'}\sigma'} 
\end{eqnarray} 
where $m_{\bold {kk'}}=\gamma_{\bold k}+\gamma_{\bold k'}$, the 
components of  the spin operator $\bold{S}_i$ are 
\begin{eqnarray} 
S_i^z&=&\frac{1}{2}(f_{i\uparrow}^{\dagger}f^{}_{i\uparrow}-f_{i\downarrow}^{\dagger}f^{}_{i\downarrow}) 
\nonumber \\ 
S_i^+&=&f_{i\uparrow}^{\dagger}f^{}_{i\downarrow} 
\nonumber \\ 
S_i^-&=&f_{i\downarrow}^{\dagger}f^{}_{i\uparrow} 
\end{eqnarray} 
and the components of $\bold {s}_{\sigma \sigma'}$ are the Pauli 
matrices divided  
by two. $t_{ij}^{(1)}$ and $t_{ij}^{(2)}$ are defined as 
\begin{eqnarray} 
t_{ij}^{(1)} 
                &=& \frac{V^2}{2} \sum_{\bold{k}} e^{i{\bf k}\cdot 
{\bf R}_{ij}}2\delta_{\bold k}  
\nonumber \\ 
t_{ij}^{(2)} 
                &=& \frac{V^2}{2} \sum_{\bold{k}} e^{i{\bf k}\cdot 
{\bf R}_{ij}}\gamma_{\bold k}  
\end{eqnarray} 
 
Finally $H_{spin}$ is obtained by means of a second canonical transformation 
which eliminates $H^{(2)}$ from $\tilde {H}$ 
\begin{equation} 
H_{spin}=e^{\hat{S}}\tilde {H}e^{-\hat{S}}, 
\label{Heff1} 
\end{equation} 
By expanding the exponentials of Eq.~\ref{Heff1}, we get an 
elimination condition that defines $S$: 
\begin{equation} 
H^{(2)}+[\hat {S},\tilde{H}]=0 
\label{cond3} 
\end{equation} 
A $S$ that satisfies this equation is 
\begin{eqnarray} 
S&=&-V^2 \sum_{i,{\bf k},{\bf k}',\sigma,\sigma'} e^{i({\bf 
k-k}')\cdot {\bf r}_i} 
\frac{m_{\bold {kk}'}}{e_{\bold k}-e_{\bold k'}} 
{\bf S}^{}_i\cdot d_{{\bf k} \sigma}^{\dagger} {\bf s}^{}_{\sigma \sigma'} 
d^{}_{{\bf k'} \sigma'} 
\nonumber \\ 
&+&\sum_{i,j,\sigma} e^{i({\bf k-k}')\cdot{\bf r}_i} 
(t_{ij}^{(1)}+t_{ij}^{(2)})  
(n^{}_{i\bar\sigma}-n^{}_{j\bar\sigma}) f_{j\sigma}^{\dagger} f^{}_{i\sigma} 
\end{eqnarray} 
Through fourth order in $V$ 
\begin{equation} 
H_{spin}=H^{(0)}+\frac{1}{2}[\hat{S},\tilde{H}^{(2)}]+\frac{1}{4} [\hat 
{W}_1,[\hat {W}_1,H^{(2)}]]  
\label{4o} 
\end{equation} 
 
The second term in Eq.~\ref{4o} reduces to 
\begin{eqnarray} 
\lefteqn{\frac{1}{2}[\hat{S},\tilde{H}^{(2)}]= 
\frac{V^4}{4}  \sum_{i,j,{\bf k},{\bf k}'}  
 e^{-i ({\bf k-k}')\cdot{\bf R}_{ij}}}\nonumber\\ 
&\times&\Bigl(\frac{2\langle n^d_{\bf k}\rangle_0 
(1-\langle n^d_{\bf k'}\rangle_0)}{e_{\bf k}-e_{\bf 
k'}} m_{\bf{kk}'}^2+\frac{(t_{ij}^{(1)}+t_{ij}^{(2)})^2}{U}\Bigr) 
{\bf S}_i\cdot {\bf S}_j 
\label{h44} 
\end{eqnarray} 
where $\langle n^d_{\bold k}\rangle_0=\langle 
\frac{1}{2}\sum_{\sigma}d_{{\bold k}\sigma}^{\dagger} d^{}_{{\bold k}\sigma} 
\rangle_0$ and $\langle \cdots\rangle_0$ means the expectation value 
relative to non-interacting conduction $d$-electrons ($V=0$). The 
first term of Eq.~\ref {h44} gives the RKKY interaction. 
 
The third term of Eq.~\ref{4o},  
$H^{(4)}=\frac{1}{4} [\hat {W}_1,[\hat {W}_1,H^{(2)}]]$, can be expressed as 
\begin{eqnarray} 
H^{(4)}&=& 
\frac{V^2}{4N^2} \sum_{i,j,\bf k,k'} e^{-i ({\bf k-k}')\cdot 
{\bf R}_{ij}} 
[3\delta_{\bf k}\delta_{\bf k'} m_{\bf kk'}  
\nonumber\\ 
&+&[\delta_{\bf k}+\delta_{\bf k'}-4m_{\bf kk'}+2n^d_{\bf 
k}(3\gamma_{\bf k}+\gamma_{{\bf k}'})] y_{\bf kk'} n_{\bf 
kk'}\gamma_{\bf k}\gamma_{\bf k'} 
\nonumber\\ 
&+&4m_{\bf kk'} y_{\bf kk'}-7 \gamma_{\bf k}^2 
\gamma_{\bf k'} +2n^d_{\bf k} \gamma_{\bf k}\gamma_{\bf 
k'}(3\gamma_{\bf k}+\gamma_{\bf k'}) 
\nonumber\\ 
&+&2\gamma_{\bf k}\gamma_{\bf k'}m_{\bf kk'}]  
{{\bf S}_i\cdot {\bf S}_j} 
\end{eqnarray} 
with 
\begin{eqnarray} 
n_{\bold {kk'}} 
                    &=&\delta_{\bold k}+\delta_{\bold k'} 
\nonumber\\ 
x_{\bold {kk'}} 
                     &=&\delta_{\bold k} + \delta_{\bold k'}- 
4(\gamma_{\bold k} +   
\gamma_{\bold k'})+2f_{\bold k}(3 \gamma_{\bold k} + \gamma_{\bold k'}) 
\nonumber\\ 
y_{\bold {kk'}} 
                     &=&\delta_{\bold k} \gamma_{\bold 
k'}+\delta_{\bold k'}\gamma_{\bold k}.  
\end{eqnarray}

\section*{B} 
 
In this appendix we  show the derivation of $H_{1\, band}^{(2)}$ 
by means of a second order canonical transformation. The  
starting Hamiltonian is:

\begin{equation} 
H=H_{1\, band}^{(0)}+H_{1\, band}^{(1)} 
\end{equation} 
where $H_{1\, band}^{(0)}$ is given by Eq. (\ref {H1bd}). From Eqs. (\ref{lot}) and  
(\ref{vijs}) we can rewrite $H^{(1)}_{1\, band}$ in the following way: 
\begin{equation} 
H^{(1)}_{1\, band} =-2 \frac {UV} {\sqrt{N}|\epsilon_f|} \sum_{i,\bold{k},\sigma}  
n^{\alpha}_{i\sigma}   
(\alpha_{i\bar{\sigma}}^\dagger \beta^{}_{\bold{k}\bar{\sigma}}+ 
\alpha_{i'\bar{\sigma}}^\dagger \beta^{}_{\bold{k}\bar{\sigma}}) 
(1+ \frac {t}{\epsilon_f}s_{\bold k}) 
\end{equation}

The canonical transformation is given by the following 
equation: 
\begin{equation} 
\tilde {H}=  e^{\hat{T}_1} H e^{-\hat{T}_1}  
\end{equation} 
The terms of order $H_{1\, band}^{(1)}$ are eliminated under 
the following condition: 
 
\begin{equation} 
H_{1\, band}^{(1)}+[\hat{T}_1,H_{1\, band}^{(0)}]=0 
\label{cond11} 
\end{equation} 
Satisfying Eq. (\ref{cond11}) is 
\begin{equation} 
\hat{T}_1= 2 \frac {UV} {\sqrt{N}|\epsilon_f|} \sum_{i,\bold{k},\sigma}  
\frac {n^{\alpha}_{i\sigma}   
(\alpha_{i\bar{\sigma}}^\dagger \beta^{}_{\bold{k}\bar{\sigma}}+ 
\alpha_{i\bar{\sigma}}^\dagger \beta^{}_{\bold{k}\bar{\sigma}})} 
{E^{+}({\bold{k}})- \tilde \epsilon_f-U} (1+ \frac {t}{\epsilon_f}s_{\bold k}) 
\end{equation} 
where $s_{\bold k}=\cos{k_x}+\cos{k_y}$. 
 
In this way we get the second order part of the transformed 
Hamiltonian $\tilde {H}$: 
\begin{equation} 
\hat {H}^{(2)}_{1\, band}  
= -4\frac {U^2V^2} {N|\epsilon_f|} \sum_{i,j, \bold{k},\sigma}  
\frac {n^{\alpha}_{i\sigma} n^{\alpha}_{j\sigma}  
(\alpha_{i\bar{\sigma}}^\dagger \alpha^{}_{j\bar{\sigma}}+\alpha_{j \bar{\sigma}}^\dagger \alpha^{}_{i\bar{\sigma}})} 
{E^{+}_{\bold{k}}- \tilde \epsilon_f-U} \tilde {t}_{ij}(\bold{k}) 
\end{equation} 
where  
\begin{eqnarray} 
\tilde {t}_{ij}(\bold{k}) = \left\{\begin{array}{ll} 
1, & {\rm for}\ i=j \\ 
\frac {t}{|\epsilon_f|} s_{\bold k}, & \mbox{for  
$i$ and $j$ n. n.} \\ 
\frac {t^2}{|\epsilon_f|^2} s_{\bold k}^2,  &\mbox{for $i$ and $j$ second and third n. n.} 
\end{array} 
\right.\nonumber\\ 
\end{eqnarray} 
Here we have considered that the lower band  
is dispersionless with $E^{-}({\bold{k}}) \sim \tilde \epsilon_f$ and  
$\bar \epsilon_f= \tau^{\beta}_{ii}-\tau^{\alpha}_{ii}$.


\end{multicols} 
 
\end{document}